%% file: main.tex
\documentclass[aps,english,twocolumn,groupedaddress]{revtex4-1}

\usepackage{graphicx}
\usepackage{amssymb}
\usepackage{amsmath}
\usepackage{mathtools}

\usepackage[colorlinks=true,
            linkcolor=red,
            citecolor=blue,
            urlcolor=blue]{hyperref}

\begin{document}

\title{Numerical study of the temperature dependence of the NMR relaxation rate\\ across the superfluid---Bose glass transition in one dimension}

\author{Maxime Dupont}
\affiliation{Department of Physics, University of California, Berkeley, California 94720, USA}
\affiliation{Materials Sciences Division, Lawrence Berkeley National Laboratory, Berkeley, California 94720, USA}

\date{\today}

\begin{abstract}
    We study the nuclear magnetic resonance (NMR) spin-lattice relaxation rate $1/T_1$ in random one-dimensional spin chains as function of the temperature and disorder strength. In the zero temperature limit, the system displays a disorder-induced quantum phase transition between a critical Tomonaga-Luttinger liquid (TLL) phase and a localized Bose glass phase. The $1/T_1$ is investigated across this transition using large-scale simulations based on matrix product state techniques. We find that this quantity can detect the transition and probe the value of the dimensionless TLL parameter $K$. We also compute the NMR relaxation rate distributions for each temperature and disorder strength considered. In particular we discuss the applicability of the stretched exponential fit to the return-to-equilibrium function in order to extract the $1/T_1$ experimentally. The results presented here should provide valuable insights in regards of future NMR experiments in realistic disordered spin compounds.
\end{abstract}

\maketitle

\section{Introduction}

Understanding phase transitions is one of the cornerstones of condensed matter physics. Among these, disorder-induced quantum phase transitions can lead to fascinating phenomena with exciting new phases of matter. A famous example is the Anderson localization in the absence of interaction~\cite{anderson1958,abrahams1979}, where the electronic wave function is spatially confined due to impurities and resulting destructive quantum interferences. Here, the random environment --- disorder --- can completely block the transport and drive a metal-to-insulator phase transition. The presence of interactions and its interplay with disorder can qualitatively change the picture and lead to the many-body localization (MBL) phenomenon. Whereas MBL commonly refers to high energy properties of interacting disordered systems~\cite{gornyi2005,basko2006,altman2015,nandkishore2015,abanin2017,alet2018}, low-energy physics is also concerned with remaining open questions.

At zero temperature, this elusive quantum many-body localized phase of matter is known as Bose glass (BG), and first appeared in the context of ${}^4$He in porous media~\cite{fisher1988,fisher1989}. While no simple microscopic picture clearly emerges, the set of properties defining this quantum phase of matter, describing a bosonic fluid lacking superfluid coherence in a random environment, is pretty well-established: It has a finite compressibility, its low-energy spectrum is gapless, correlations are short-ranged, and there is no global phase coherence. Besides helium-4 in random media, the phase has been reported in various experimental systems such as amorphous indium oxide films with a transition from a superconducting to insulator phase~\cite{shahar1992,sacepe2008,sacepe2011}. It has also been observed in an array of quasi-one-dimensional samples of ${}^{39}$K cold atoms, subject to a quasiperiodic optical lattice~\cite{derrico2014}. Another type of systems in which the Bose glass phase has been investigated are antiferromagnetic Mott insulators~\cite{nohadani2005,hida2006,roscilde2006,zheludev2013} such as the spin-$1/2$ ladder compound $($CH$_3)_2$CHNH$_3$Cu$($Cl$_x$Br$_{1-x})_3$~\cite{hong2010} and the Br doped spin-$1$ system Ni(Cl$_{1-x}$Br$_x$)$_2$-4SC(NH$_2$)$_2$~\cite{yu2010,yu2012,yu2012prb,wulf2013,orlova2017,dupont2017,dupont2017prb,orlova2018} to cite but a few~\cite{fumiko2011,kamiezniarz2016}.

In one dimension, the microscopic mechanisms driving the superfluid-to-Bose-glass transition are still controversial~\cite{altman2004,altman2008,altman2010,ristivojevic2012,pollet2013,pielawa2013,pollet2014,ristivojevic2014,yao2016}. Even in absence of disorder, in one dimensional systems, quantum fluctuations prevent the emergence of a global phase coherence that would result from the spontaneous breaking of a continuous symmetry. At best, one can expect a critical Tomonaga Luttinger Liquid (TLL) phase with a finite superfluid density and quasi-long-range order characterized by power-law decaying transverse correlations $\propto r^{-1/2K}$ at large distance $r$, where $K$ is the so-called TLL parameter~\cite{giamarchi2004}. In the presence of disorder, the TLL phase is expected to be unstable towards a Bose glass phase, unless $K>3/2$. In the latter case, a critical disorder strength is required to drive the system from a superfluid to a many-body-localized phase. Giamarchi and Schulz have shown in their seminal work that this transition belongs to the Berezinskii-Kosterlitz-Thouless (BKT) universality class~\cite{berezinskii1971,kosterlitz1973,kosterlitz1974} with $K\equiv K_c$ taking a finite value at criticality~\cite{giamarchi1987,giamarchi1988}. The TLL parameter might take the universal value $K_c=3/2$ at weak disorder (compared to the bandwidth)~\cite{ristivojevic2012,ristivojevic2014} but the extension of this weak disorder regime remains unclear and other approaches~\cite{altman2004,altman2008,altman2010,pollet2013,pielawa2013,pollet2014,yao2016} based on strong disorder analysis suggest a non-universal value. Only recently a numerical study attempted to clarify the different scenarios proposed and to precisely define the weak versus strong disorder regimes~\cite{doggen2017}.

Spin compounds are one of the best candidates to address these open issues experimentally. For instance, the nuclear magnetic resonance (NMR) spin-lattice relaxation rate $1/T_1$ has proven to be a formidable probe for one-dimensional physics in Mott insulators~\cite{berthier2017} and might be very well-suited for the purpose. The $1/T_1$ in a TLL phase is expected to diverge algebraically $\propto T^{1/2K-1}$ at low temperature~\cite{sachdev1994,chitra1997,barzykin2001,klanjsek2008}. This prediction has been perfectly checked against numerics in paradigmatic (clean) $S=1/2$ XXZ spin chains~\cite{coira2016,dupont2016} for $T/J\lesssim 10$, where $J$ is the antiferromagnetic exchange coupling. Moreover, the parameter $K$ has been reliably extracted by fitting experimental $1/T_1$ measurements versus $T$ in various quasi-one-dimensional spin compounds, such as the spin$-1/2$ Heisenberg antiferromagnetic ladder $($C$_7$H$_{10}$N$)_2$CuBr$_4$ (DIMPY)~\cite{jeong2013,jeong2016,moller2017}. The effect of inherent (albeit weak) three-dimensional couplings $J_\mathrm{3D}$ on the $1/T_1$ in spin materials has also been recently studied in Ref.~\onlinecite{dupont2018} where the authors could properly define a temperature window $J_\mathrm{3D}/100\lesssim T\lesssim J/10$ for observing genuine one-dimensional physics. Therefore, this quantity might provide a final experimental answer to the disputed issue of $K_c$. Indeed, strong-leg ladders subject to an external magnetic field can realize a TLL with $K>3/2$~\cite{giamarchi1999,hikihara2001,giamarchi2004} and numerous examples of observation of the BG phase have been given above, making us confident that, eventually, an ideal spin compound combining a strong enough attractive regime and controllable chemical disorder will be experimentally available in the future.

In this work, using large-scale simulations based on matrix product state, we study the $1/T_1$ across the superfluid—Bose glass transition in paradigmatic one-dimensional XXZ spin$-1/2$ chains with a negative Ising anisotropy and subject to a random magnetic field, as a function of the temperature and disorder strength. The results presented here should provide valuable insights in regards of future NMR experiments.

The rest of the paper is organized as follows. In Sec.~\ref{sec:model_defs}, we provide an overview of the studied model, the definition of the NMR spin-lattice relaxation rate $1/T_1$ and briefly present the numerical techniques. Results are then discussed in Sec.~\ref{sec:results}. Finally, we summarize our conclusions in Sec.~\ref{sec:summary_conclusions}.

\section{Model, definitions and methods}\label{sec:model_defs}

\subsection{Model}

We consider the spin-$1/2$ XXZ chain in a random magnetic field described by the following Hamiltonian:
\begin{equation}
    \mathcal{H}=J\sum_i\Biggl[\frac{1}{2}\Bigl(S^+_iS^-_{i+1}+\mathrm{H.c.}\Bigr)+\Delta S^z_iS^z_{i+1} + h_iS^z_i\Biggr],
    \label{eq:hamiltonian}
\end{equation}
with $i$ labeling the lattice sites, $J$ the overall energy coupling, set to unity in the following, and $\Delta\in(-1,1]$ is the Ising anisotropy. The random variables $h_i$ are drawn independently from a uniform distribution $\in[-h, h]$ where $h$ characterizes the disorder strength.

\begin{figure}[!t]
    \center
    \includegraphics[width=0.7\columnwidth,clip]{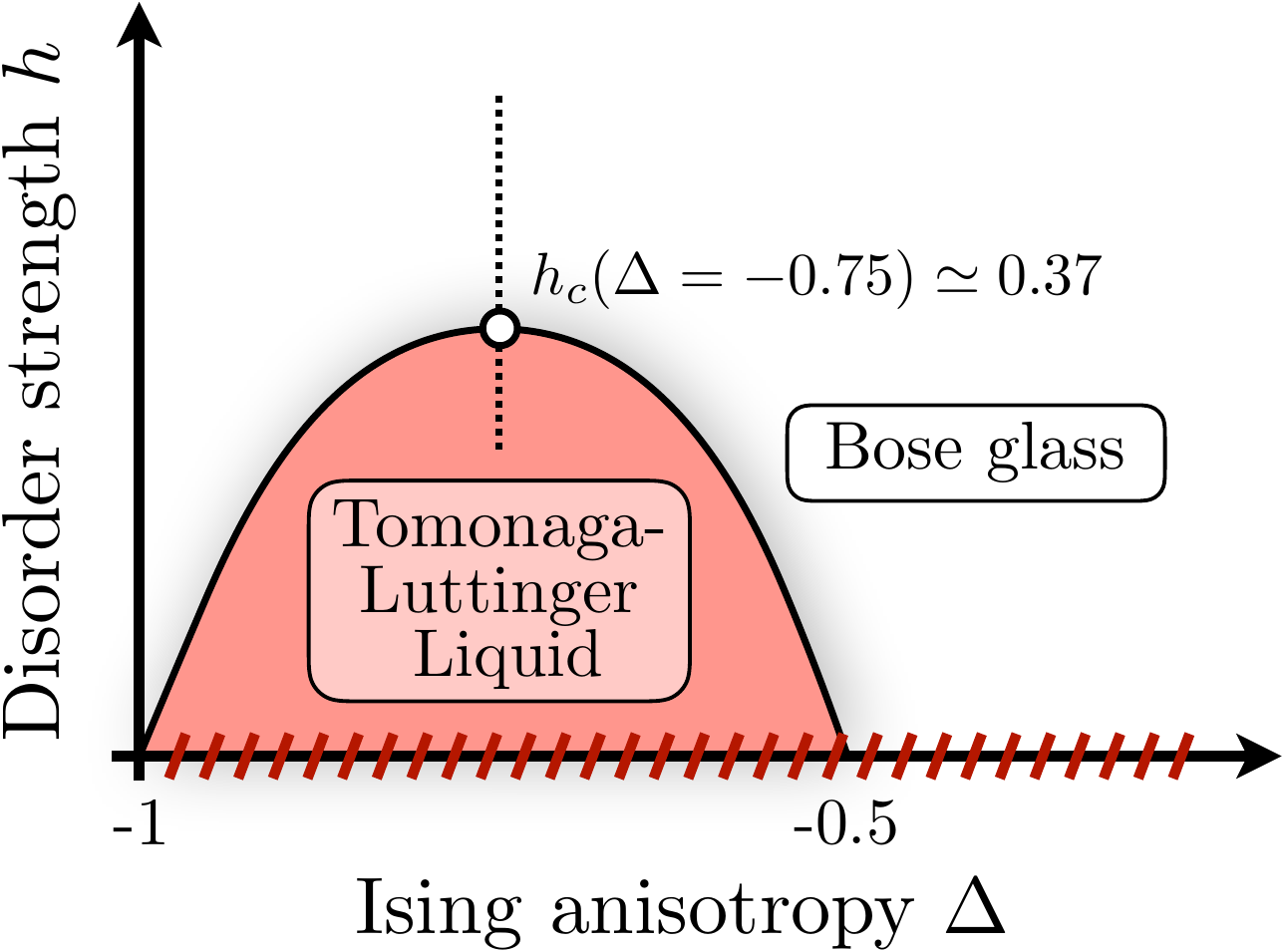}
    \caption{Schematic zero temperature ``disorder strength $h$ versus Ising anisotropy $\Delta$'' phase diagram of the Hamiltonian~\eqref{eq:hamiltonian}. For $h=0$ (no disorder) the model is in a critical Tomonaga-Luttinger Liquid phase for $\Delta\in (-1,1]$ (red stripes). From this line, any amount of disorder drives the system to a localized Bose glass phase for $\Delta>-0.5$. For smaller Ising anisotropies, a finite amount of disorder is required to induce a quantum phase transition (plain black line). The maximum of the dome is around $\Delta=-0.75$ with a critical disorder strength $h_c\simeq 0.37$~\cite{doggen2017}. This work focuses on the $\Delta=-0.75$ line of the phase diagram (dotted black line) for various disorder strengths and temperatures.}
    \label{fig:phase_diagram}
\end{figure}

In the clean ($h=0$) case, it is well-known that in the low-energy limit, the model~\eqref{eq:hamiltonian} can be described as a Tomonaga-Luttinger liquid~\cite{giamarchi2004} with algebraically decaying correlations and a finite superfluid density (spin stiffness). This description only relies on two phenomenological parameters: $u$, the propagation velocity of the excitations in the system and $K$, the dimensionless TLL parameter governing for instance the decay of correlations. Here, $u$ and $K$ can be related to the microscopic parameter $\Delta$ from Bethe ansatz equations~\cite{luther1975}. Through a Matsubara-Matsuda transformation~\cite{matsubara1956}, the spin Hamiltonian~\eqref{eq:hamiltonian} can be mapped exactly to a hard-core bosons (HCB) model with $\Delta$ controlling the nearest-neighbor HCB interaction. Precisely, $\Delta<0$ ($K>1$) corresponds to an attractive interaction between the bosons and this part of the phase diagram is often referred to as the \textit{attractive regime}. In presence of disorder ($h>0$), the critical TLL phase is expected to be unstable towards a localized Bose glass phase with exponentially decaying correlations and a vanishing superfluid density. However, for strong enough attractive interaction, $K>3/2$ ($\Delta<-0.5$), a finite disorder strength $h_c$ is necessary to drive the system from a TLL to a Bose glass phase, with the TLL parameter $K_c(\Delta,h_c)$ taking a finite value at the transition~\cite{giamarchi1987,giamarchi1988}. A schematic phase diagram of model~\eqref{eq:hamiltonian} is shown in Fig.~\ref{fig:phase_diagram}.

In this work, we focus on $\Delta=-0.75$ and vary the disorder strength $h$ from zero to $1/2$, covering the TLL and Bose glass phases with the transition happening at $h_c\simeq 0.37$~\cite{doggen2017}. Importantly, the temperature will be a parameter to be as close as possible to realistic experimental setups. Although the model~\eqref{eq:hamiltonian} is unrealistic to describe spin materials because of the random fields with no reasonable origin, the essential ingredients are: (i) in the low-energy limit and in the clean case, a TLL description of the spin compound with a TLL parameter $K>3/2$, (ii) disorder whose microscopic origin is likely compound-dependent but which drives the system to a localized phase if sufficiently strong.

\subsection{The NMR spin-lattice relaxation rate $1/T_1$}

In experiments, the NMR spin-lattice relaxation rate $1/T_1$ is the inverse characteristic time that the targeted nuclear spins, polarized through a static external magnetic field, take to relax back to their thermodynamic equilibrium after a perturbation by an electromagnetic pulse of frequency $\omega_0$, changing their polarization~\cite{abragam1961,horvatic2002,slichter2013}. The return-to-equilibrium process can be described as function of time $t$ as
\begin{equation}
    1-M(t)\propto\mathrm{e}^{-t/T_1},
    \label{eq:return_to_eq}
\end{equation}
where $M(t)$ is the component of the nuclear spins along the static magnetic field. One can show that $1/T_1$ provides information on the electronic spins $\mathbf{S}_i=(S^x_i,S^y_i,S^z_i)$, i.e., those described by the Hamiltonian~\eqref{eq:hamiltonian}, through the following expression,
\begin{equation}
    \frac{1}{T_1}=\frac{\gamma}{2}\sum_{a,b\in[x,y,z]}A_{ab}^2\int_{-\infty}^{+\infty}\mathrm{d}t\,\mathrm{e}^{it\omega_0} C_{a,b}(t),
    \label{eq:t1_def1}
\end{equation}
with $\gamma$ the gyromagnetic ratio, $A_{ab}$ the hyperfine coupling constant describing the interaction between nuclear and electronic spins and $C_{a,b}(t)$ the dynamical correlation function
\begin{equation}
    C_{a,b}(t) = \langle S^a_i(t)S^b_i(0)\rangle-\langle S^a_i(t)\rangle\langle S^b_i(0)\rangle,
    \label{eq:t1_corr}
\end{equation}
which is a local quantity because measured on the same site $i$ for both operators. Here $\langle\rangle$ indicates the thermal average and $S^a_i(t)=\mathrm{e}^{i\mathcal{H}t}S^a_i\mathrm{e}^{-i\mathcal{H}t}$ is the time-dependent spin operator in the Heisenberg picture. It is theoretically justified to take the limit $\omega_0=0$ since the NMR frequency is usually the smallest energy scale of the problem~\footnote{Taking this limit assumes some smoothness in the Fourier transform of $C_{a,b}(t)$ (time $t$ to frequency $\omega$), with no sharp contribution at $\omega\rightarrow 0$ that would not be captured by the NMR measurements because $\omega_0$ is actually finite. This is the case for the model studied here.}. Moreover, due to the $\mathrm{U}(1)$ symmetry of the Hamiltonian~\eqref{eq:hamiltonian}, the $1/T_1$ can be reduced to the components $(a,b)=(\pm,\mp)$ and $(z,z)$, where the $x$ and $y$ spin components have been expressed using the raising and lowering operators~\footnote{In the $\omega_0\rightarrow 0$ limit, the contribution of $(+,-)$ becomes equivalent to $(-,+)$ and one can choose to work with the most convenient to compute their contribution to the NMR spin-lattice relaxation rate $1/T_1$.}. The definition~\eqref{eq:t1_def1} now reads
\begin{equation}
    \frac{1}{T_1}=\gamma\int^{+\infty}_0\mathrm{d}t\,\Biggl\{A^2_{\pm\mp} \operatorname{Re}\Bigl[C_{\pm,\mp}(t)\Bigr] + A^2_{zz} \operatorname{Re}\Bigl[C_{z,z}(t)\Bigr]\Biggr\},
    \label{eq:t1_def2}
\end{equation}
where the first and second terms of Eq.~\eqref{eq:t1_def2} can be labeled as transverse and longitudinal contributions respectively. In the following, we only focus on the transverse contribution which dominates over the longitudinal one from intermediate to low temperatures and set the experiment-dependent prefactor $\gamma A^2_{\pm\mp}$ to unity. For a TLL, the  transverse contribution to the NMR relaxation rate has been found for $\omega_0/T\ll 1$ in the form~\cite{sachdev1994,chitra1997,barzykin2001,klanjsek2008}
\begin{equation}
    \frac{1}{T_1} = \frac{A\cos\left(\frac{\pi}{4K}\right)\mathrm{B}\left(\frac{1}{4K},1-\frac{1}{2K}\right)}{u}\left(\frac{2\pi T}{u}\right)^{\frac{1}{2K}-1},
    \label{eq:t1_tll}
\end{equation}
with $\mathrm{B}(x,y)$ the Euler beta function, $u$ and $K$ the TLL parameters and $A$ the prefactor of the static correlation function $\langle S^\pm_r(0)S^\mp_0(0)\rangle$ at zero temperature. The prediction~\eqref{eq:t1_tll} perfectly checks against numerics at low temperature, $T/J\lesssim 10$, without any adjustable parameter~\cite{coira2016,dupont2016}.

\subsection{Numerical methods}

\subsubsection{Finite temperature}

To compute the local dynamical correlation function~\eqref{eq:t1_corr} at finite temperature, we use the Matrix Product State (MPS) formalism~\cite{schollwock2011}. Whereas a MPS represents a pure state, it can also be used for mixed state through the purification method~\cite{verstraete2004}. The basic idea is to write the density matrix as a pure state in an enlarged Hilbert space with half physical and half auxiliary degrees of freedom (they can be taken as a copy of the physical ones). From a practical point of view, the corresponding infinite temperature pure state can be written down exactly as a MPS of bond dimension $m=1$: A product state of maximally entangled pairs of physical and auxiliary degrees of freedom. One can show that the pure state at inverse temperature $\beta=1/T$ is obtained by time-evolving the infinite temperature one with $\mathrm{exp}{(-\beta\mathcal{H}/2)}$, where $\mathcal{H}$ only acts on physical degrees of freedom. We perform the imaginary-time evolution using the time-evolving block decimation (TEBD) algorithm~\cite{vidal2004} along with a fourth order Trotter decomposition~\cite{hatano2005} and time-step $\delta_\beta=0.1$.

\subsubsection{Real-time evolution}

When the desired finite temperature state is obtained, a real-time evolution with $\mathrm{exp}{(-i\mathcal{H}t)}$ is carried out using the same TEBD algorithm as for the imaginary-time evolution (fourth order and time-step $\delta_t=0.1$). This is the most limiting part since the real time evolution of a quantum state produces a rapid growth of entanglement entropy~\cite{laflorencie2016} while the efficiency of the MPS representation relies on low entangled states. In practice, this limits the maximum time $t$ one can reach in the simulation. Some workarounds have been developed in order to push the limits further like the linear prediction~\cite{barthel2013} or evolving the auxiliary degrees of freedom with $-\mathcal{H}$ in real time~\cite{karrasch2013}. We have used the latter, which can be seen as a local disentangling operation, of which the real time evolution by $-\mathcal{H}$ is just one possibility~\cite{hauschild2018}. In this work, the maximum bond dimension of the MPS was set to $m=500$. Furthermore, the NMR relaxation rate being a local quantity, the dynamical correlation~\eqref{eq:t1_corr} is computed in the middle of the chain to avoid open boundary and finite size effects. Indeed, at long time and at finite temperature, the correlation decays exponentially over time with a characteristic time $\tau(T,h)$ which makes the choice of a sufficiently large system $L(T,h)$ enough to consider the results in the thermodynamic limit, i.e., $u(T,h)\tau(T,h)\ll L(T,h)$ with $u(T,h)$ some velocity accounting for the spreading of the excitation $S_i^{\pm}$ in the chain. This is especially true in regards of the definition~\eqref{eq:t1_def2} for the NMR relaxation rate as an integral over time of this exponentially decaying correlation function. This integration is performed numerically using the standard Simpson's rule.

\subsubsection{Disorder sampling}

In the following, the local dynamical correlation~\eqref{eq:t1_corr} is computed over $N_\mathrm{s}\approx 500$ independent samples for each temperature and disorder strength considered. Therefore, it is the most demanding part numerically since the simulation of each independent sample is already quite demanding itself.

\section{Results}\label{sec:results}

\subsection{Time dependence of the local spin-spin correlation function}

We first look at the real part of the transverse dynamical correlation function $C_{\pm,\mp}(t)$ defined in Eq.~\eqref{eq:t1_corr} and whose integral over time gives the NMR spin-lattice relaxation rate $1/T_1$ according to Eq.~\eqref{eq:t1_def2}.

It is instructive to consider the disorder-free $h=0$ and $\Delta=0$ case, which in the low-energy limit, also belongs to the TLL phase. At this specific point of the phase diagram, an expression for $C_{\pm,\mp}(t)$ can be derived exactly and expressed as a Pfaffian~\cite{stolze1992,stolze1995}. In some cases, it can be brought into a more explicit form, and one finds three distinct regions in time. After a very short time of order $\sim\mathcal{O}(1)$ (region A), the correlation shows a power-law decay (region B) before displaying an exponential decay $\propto \mathrm{exp}[-t/\tau(T)]$ at longer time (region C), where the decay time diverges algebraically with temperature $\tau(T)\propto 1/T$ for $T\ll J$. It is also well-known that the thermal correlation length $\xi(T)$ diverges at low temperature in the same way, $\propto u/T$ with $u$ the TLL velocity~\cite{pereira2012}, so that in the end, one can relate those characteristic thermal time and space quantities through $u\tau(T)\sim\xi(T)$. The intermediate power-law region grows larger and larger as the temperature is reduced to eventually take over the long time exponential decay at zero temperature, with no characteristic length scale in this limit, the system being critical. We note $\tau_*(T)$ the crossover time between regions B and C.

Away from the $\Delta=0$ point, no exact expression is available, but no major qualitative difference is expected at low temperature, with a finite thermal correlation length $\xi(T)\propto u/T$ and the existence of a characteristic thermal time scale $\tau(T)$: It is still expected to follow $\propto 1/T$~\cite{pereira2012}, which we verified numerically at $h=0$ for $\beta J\gtrsim 5$ (data not shown). At high temperature however, and as pointed out in Ref.~\onlinecite{stolze1995}, it might not be as simple to relate the two quantities, displaying different temperature dependences: For instance at infinite temperature $\xi(\infty)$ is zero while $\tau(\infty)$ remains finite. In any case there exists, for the transverse dynamical correlation function in the TLL regime, characteristic thermal length and time scales in the temperature window considered in this work. More generally, they also depend on the disorder strength, i.e., $\xi(T,h)$, $\tau(T,h)$, $\tau_*(T,h)$, and correspond in that case to the disorder averaged quantity.

\begin{figure}[!t]
    \center
    \includegraphics[width=1\columnwidth,clip]{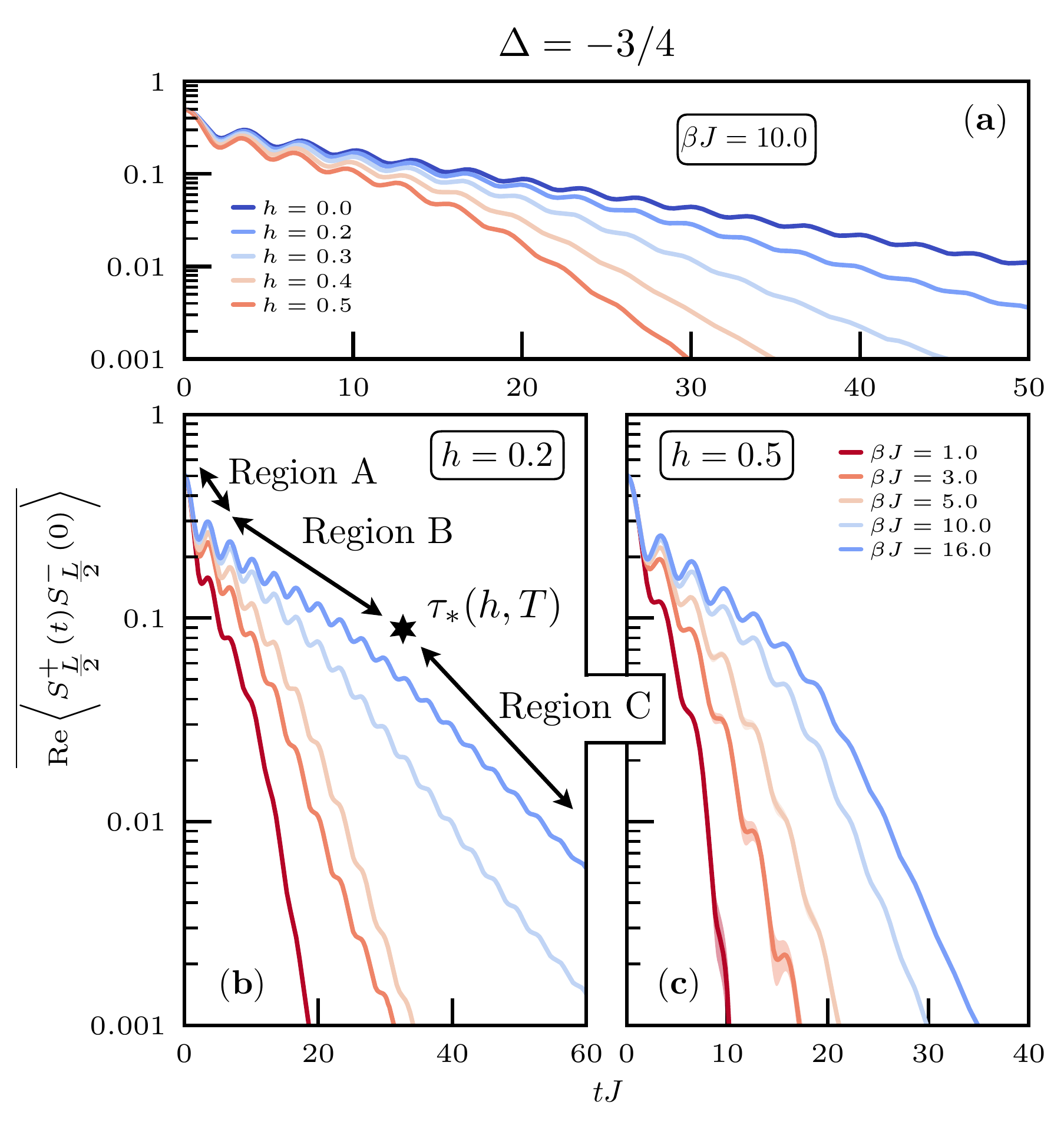}
    \caption{Average value of the real part of the dynamical correlation function $\langle S^+_{L/2}(t)S^-_{L/2}(0)\rangle$ versus time for various values of disorder strengths $h$ at fixed inverse temperature $\beta J=10.0$ (panel~\textbf{a}) and for various inverse temperatures $\beta J$ at fixed disorder strengths $h=0.2$ (panel~\textbf{b}) and $h=0.5$ (panel~\textbf{c}). The average is performed over $N_\mathrm{s}\approx 500$ independent samples, except for $\beta J\geq 5$ at $h=0.2$ where only $N_\mathrm{s}\approx 200$ samples are available. The legend is the same for panels~\textbf{b} and~\textbf{c}. The dynamical correlation is measured in the middle of open chains of length $L(T,h)$. The integral of this quantity gives the NMR spin-lattice relaxation rate $1/T_1$ according to Eq.~\eqref{eq:t1_def2}. After a very short time $tJ\sim\mathcal{O}(1)$ (region A), the average real part of the dynamical correlation displays an intermediate power-law regime (region B) before undergoing an exponential suppression over time (region C). The crossover time between regions B and C is $\tau_*(T,h)$ and the decay time of region C is $\tau(T,h)$.}
    \label{fig:real_part_vs_time}
\end{figure}

\begin{figure*}[!ht]
    \includegraphics[width=2\columnwidth,clip]{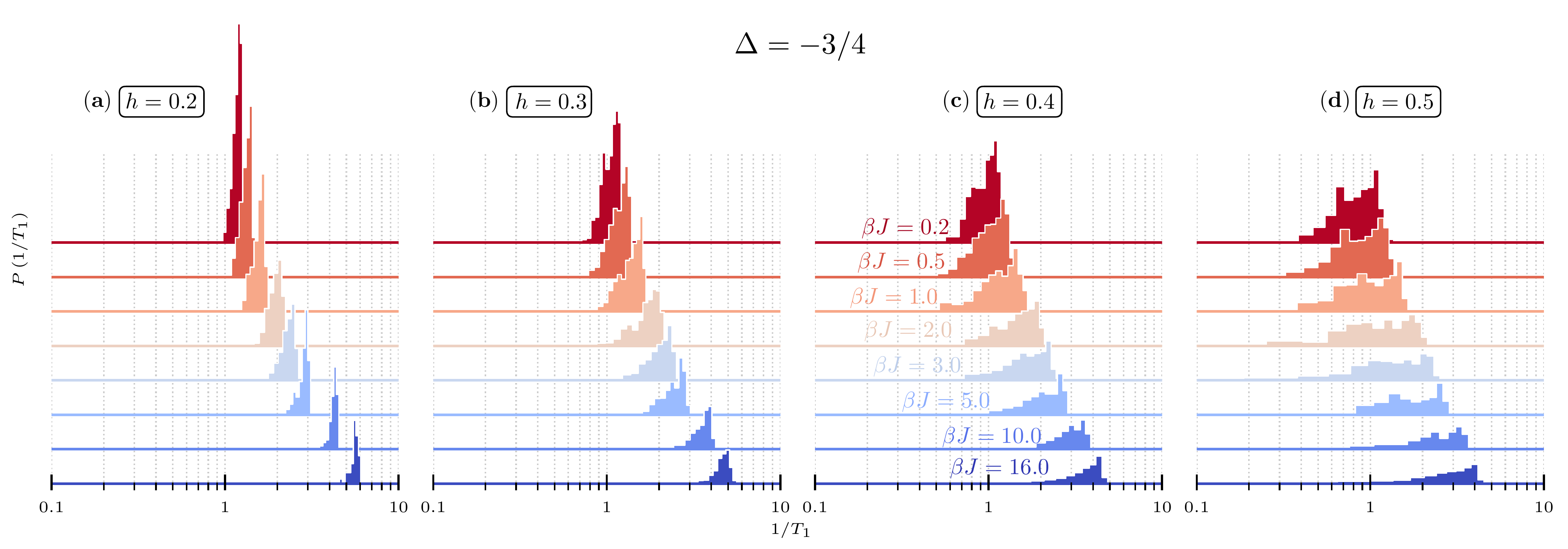}
    \caption{Probability distribution $P(1/T_1)$ of the NMR spin-lattice relaxation rate for various disorder strengths: (\textbf{a}) $h=0.2$, (\textbf{b}) $h=0.3$, (\textbf{c}) $h=0.4$ and (\textbf{d}) $h=0.5$. Note the log scale for the $x$ axis. The different colors correspond to different inverse temperatures $\beta J$, as specified in~(\textbf{c}). The data have been shifted vertically for visibility. At zero temperature, the disorder-induced quantum phase transition between the TLL and Bose glass phases happens at $h_c\simeq 0.37$~\cite{doggen2017}, see Fig.~\ref{fig:phase_diagram}. Each distribution is made of $N_\mathrm{s}\approx 500$ independent disordered samples, except for $h=0.2$ and $\beta J\geq 5$ with only $N_\mathrm{s}\approx 200$ available.}
    \label{fig:distributions}
\end{figure*}

Plugging in disorder, the position of the crossover time scale $\tau_*(T,h)$ becomes more manifest with the decay time $\tau(T,h)$ being shorter as the disorder strength and temperature are increased. This results in a steepest slope of the long-time exponential decay, as visible in all three panels of Fig.~\ref{fig:real_part_vs_time}. By increasing these parameters, one expects quantum coherence to be generally weaker over time, hence the shorter characteristic and crossover times $\tau(T,h)$ and $\tau_*(T,h)$. In the localized regime shown in Fig.~\ref{fig:real_part_vs_time}\,(c) for $h=0.5$, the width of the intermediate power-law region should be bounded, not only by thermal, but also by localization effects. Indeed, taken separately, both effects induce a finite length scale in the system: A thermal length $\xi_\mathrm{th}(T,h)$ or a localization length $\xi_\mathrm{loc}(T=0,h\neq 0)$. The dominant effect will be associated to the smallest of the two length scales. Deep in the localized regime, the disorder averaged localization length follows $\xi_\mathrm{loc}(T=0,h\neq 0)\propto h^{2/(2K-3)}$~\cite{giamarchi1987,giamarchi1988} and it diverges exponentially when approaching the transition from the localized phase $h\rightarrow h_c$, expected behavior for a BKT transition~\cite{berezinskii1971,kosterlitz1973,kosterlitz1974}.

Finally, the NMR spin-lattice relaxation rate is the integral over time of the real part of the dynamical correlation function displayed in Fig.~\ref{fig:real_part_vs_time}. Because it decays exponentially with time, reaching a finite maximum time $t_\mathrm{max}$ is actually sufficient numerically to get an accurate estimate of the $1/T_1$.

\subsection{NMR spin-lattice relaxation rate distributions}

Computing the dynamical correlation~\eqref{eq:t1_corr} for independent disorder samples $(N_\mathrm{s}\approx 500)$ allows us to establish the corresponding distribution fo the NMR relaxation rate $P(1/T_1)$ for each temperature and disorder strength, covering both TLL and BG phases as shown in Fig.~\ref{fig:distributions}. Overall, it is at high temperature that the distributions seem to be the narrower, independently of the disorder strength. With a very short thermal correlation length $\xi(T/J\gg 1,h)\ll 1$, one expects the effect of temperature to be dominant over disorder surrounding the site $i$ on which the local dynamical correlation function is being computed.

Interestingly, in the BG phase, see Fig.~\ref{fig:distributions}\,(d), the $1/T_1$ distributions for $\beta J\lesssim 10$ seem to be double-peaked. This can be understood as a competition between thermal and localization effects. At fixed temperature and for a given disordered sample, the relevant characteristic length $\xi(h,T)$ surrounding the local site from which the $1/T_1$ is computed, is either going to be $\xi_\mathrm{th}$ or $\xi_\mathrm{loc}$, whichever is the shortest one. The distribution of $\xi(h,T)$ over many samples is going to be a weighted combination of the respective distributions of thermal and localization lengths, leading to a bimodal structure of the overall distribution. As discussed previously, a finite length scale induces a finite time scale, either through the decay time $\tau(T,h)$ or the crossover time $\tau_*(T,h)$, in the local dynamical correlation function, which will be reflected on the $1/T_1$ distributions by also displaying a double-peaks structure. In Fig.~\ref{fig:distributions}\,(d) for $\beta J =16$, we do not observe two modes anymore because localization is dominant at that temperature. For $h=0.4$, see Fig.~\ref{fig:distributions}\,(c), no double-peaks structure clearly emerges despite being in the localized BG phase. This is because we are very close to the transition, $h_c\simeq 0.37$~\cite{doggen2017}, with the localization length diverging exponentially, making it way larger than the thermal length.

Experimentally, one does not have access to the $1/T_1$ distributions but, for a disordered system to the average value of the return-to-equilibrium function $M(t)$ of Eq.~\eqref{eq:return_to_eq},
\begin{equation}
    1-\overline{M(t)}\propto\overline{\mathrm{e}^{-t/T_1}}=\int_0^{+\infty}\mathrm{d}T_1^{-1}\,P\left(T_1^{-1}\right)\mathrm{e}^{-t/T_1},
    \label{eq:return_to_eq_dis}
\end{equation}
from which the average NMR spin-lattice relaxation rate $\overline{1/T_1}$ cannot be readily obtained. In order to access it, one usually makes the assumption that the sum of $\mathrm{exp}(-t/T_1)$ originating from the disorder averaging in Eq.~\eqref{eq:return_to_eq_dis} can be approximated by a stretched exponential~\cite{johnston2006},
\begin{equation}
    \overline{\mathrm{exp}(-t/T_1)}\simeq\mathrm{exp}\Bigl[-(t/\tau_\mathrm{str})^\theta\Bigr],
    \label{eq:str_exp}
\end{equation}
with $\theta$ and $\tau_\mathrm{str}$ two parameters fitted against the experimental data. One then considers that $\tau_\mathrm{str}\equiv T_1$, which is exact in the disorder-free case where $\theta=1$. Under the stretched exponential assumption, the distribution $P(1/T_1)$ corresponds to the stretched exponential one~\cite{johnston2006}. It is clear that the distributions displayed in Fig.~\ref{fig:distributions} do not correspond to pure stretched exponential distributions defined on the semi-infinite interval $[0,+\infty]$. For the disorder strengths considered, we do not observe rare events in the $1/T_1$ value leading to very broad distributions, over orders of magnitudes, as it was for instance observed in spin chains with random exchange couplings~\cite{shu2018}. Here, the NMR spin-lattice relaxation rate seems to be bounded by its clean value at $h=0$. Moreover the stretched exponential distribution would not capture the double peaks at high temperature in the BG phase. Yet, surprisingly, the value of $1/\tau_\mathrm{str}$ extracted from the approximation~\eqref{eq:str_exp} is very close to the average $\overline{1/T_1}$ value, as discussed in the following.

\subsection{Temperature dependence of the NMR spin-lattice relaxation rate $1/T_1$}

\begin{figure}[!t]
    \center
    \includegraphics[width=1\columnwidth,clip]{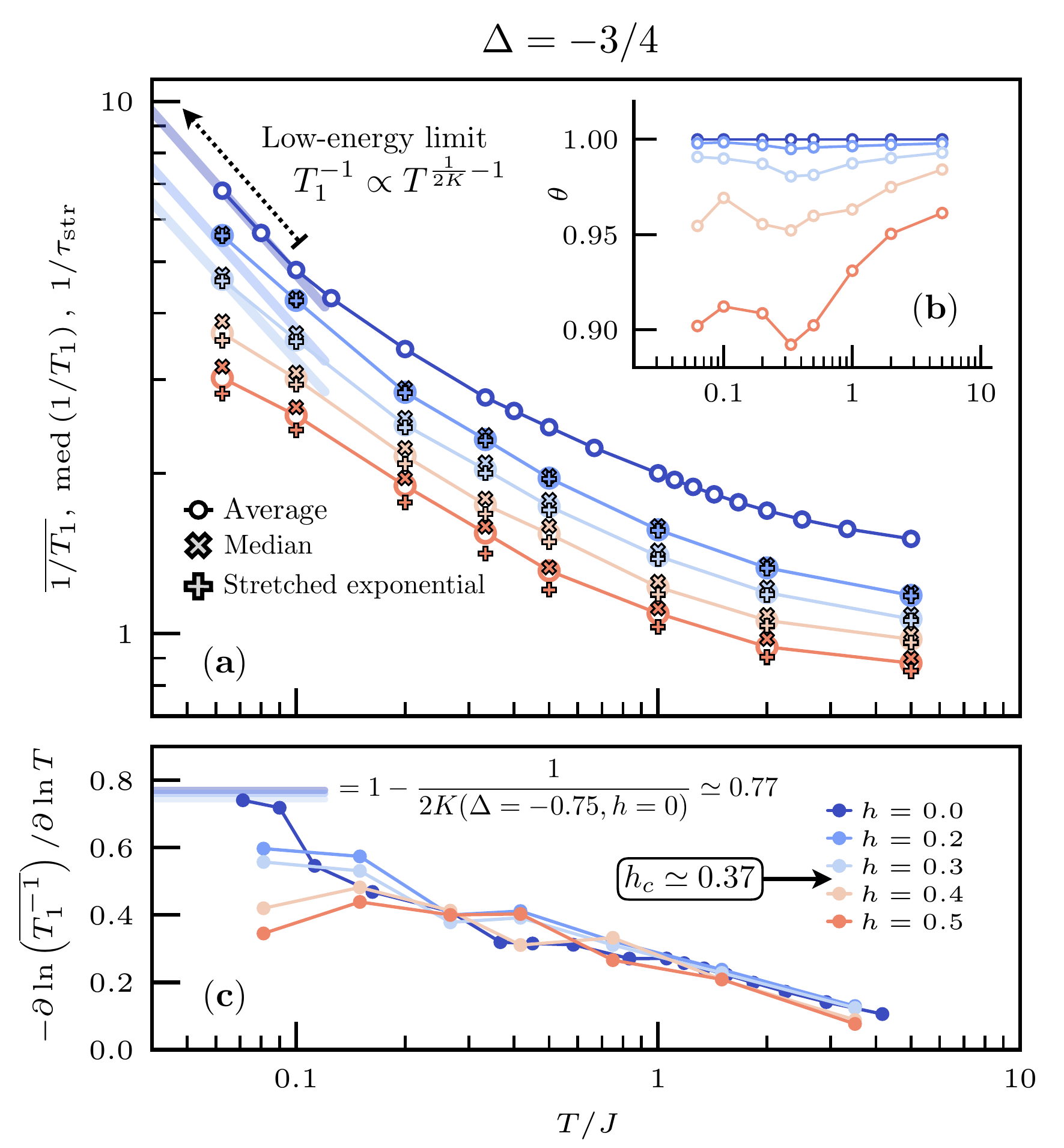}
    \caption{Legend displayed in panel (\textbf{c}). (\textbf{a}) Log-log scale. Temperature dependence of the average (circle symbol), median (cross symbol) and stretched exponential (plus symbol) values of the NMR spin-lattice relaxation rate, versus temperature $T/J$ for various disorder strengths $h=0.0$ (clean case, no disorder), $0.2$, $0.3$, $0.4$, and $0.5$. At zero temperature, the disorder-induced quantum phase transition between the TLL and Bose glass phases happens at $h_c\simeq 0.37$~\cite{doggen2017}, see Fig.~\ref{fig:phase_diagram}. The purely algebraic temperature dependence of the TLL prediction, valid in the ``low-energy limit'', is displayed by the bold straight line according to Eq.~\eqref{eq:t1_tll} and the parameter shown in Tab.~\ref{tab:tll_parameters}. (\textbf{b}) Stretched exponential exponent $\theta$ versus the temperature. (\textbf{c}) Temperature dependence of the gradient of the average value, $-\partial\ln(\overline{T_1^{-1}})/\partial\ln T$, versus temperature $T/J$. The gradient is computed numerically. The transition between the TLL and Bose glass phases is marked by a change in the slope of the gradient. In the TLL regime $T_1^{-1}\propto T^{1/2K-1}$ with $K$ the TLL parameter, this quantity should saturate to $1-1/2K$. This is verified for the clean case with $1-1/2K(\Delta=-0.75,h=0)\simeq 0.77$. In presence of disorder ($h=0.2$, $0.3$) the necessity to go to lower temperatures to observe genuine TLL physics makes the gradient not converged yet to the expected value of $K$ reported in Tab.~\ref{tab:tll_parameters}.}
    \label{fig:t1_vs_temp}
\end{figure}

In the TLL regime, the NMR spin-lattice relaxation rate $1/T_1$ takes the form~\eqref{eq:t1_tll} in the low temperature limit, with \textit{no free parameter}. This expression has been perfectly checked against numerics in clean XXZ chains for various $\Delta$ anisotropies and for $T/J\lesssim 10$~\cite{coira2016,dupont2016}. For completeness, we display the disorder-free case at $\Delta=-0.75$ in Fig.~\ref{fig:t1_vs_temp}\,(a), where the crossover towards genuine the TLL physics with $1/T_1\propto T^{1/2K-1}$ is indeed recovered at low temperature. In presence of disorder, but below the critical disorder strength $h_c\simeq 0.37$~\cite{doggen2017}, we still expect a TLL phase and the algebraic temperature dependent expression in Eq.~\eqref{eq:t1_tll} should still apply. However for $h\neq 0$, there is no exact expression for the TLL parameters $K$, $u$ or the prefactor $A$ of the static transverse correlation function at zero temperature, and they must be computed numerically. These values are reported in Tab.~\ref{tab:tll_parameters} for $h=0$, $0.2$ and $0.3$. The technical details for their determination are explained in the appendix.

\begin{table}[!t]
    \center
    \begin{ruledtabular}
        \begin{tabular}{lccc}
            $h$ & $A$~(DMRG) & $u/J$~(QMC) & $K$~(QMC, DMRG) \\\hline
            $0.0$ (exact) & $0.159572$ & $0.429535$ & $2.173408$ \\
            $0.2$ & 0.13084(4) & 0.419(2) & 2.094(8),~2.0899(6) \\
            $0.3$ & 0.12448(7) & 0.404(2) & 1.958(9),~1.953(2)
        \end{tabular}
    \end{ruledtabular}
    \caption{Tomonaga-Luttinger liquid parameters $u$ and $K$ as well as the prefactor $A$ of the static transverse correlation function $\langle S^\pm_r(0)S^\mp_0(0)\rangle$ at zero temperature. For the first line, corresponding to the clean case, the parameters are known exactly~\cite{luther1975,lukyanov2003,hikihara2004}. In presence of disorder, for $h=0.2$ and $h=0.3$, the value of the parameters is determined numerically using density-matrix renormalization group (DMRG) and quantum Monte Carlo (QMC). See the appendix for technical details.}
    \label{tab:tll_parameters}
\end{table}

We first focus on the average $\overline{1/T_1}$ value, which for $\beta J\gtrsim 16$ agrees with the TLL prediction combining Eq.~\eqref{eq:t1_tll} and the parameters of Tab.~\ref{tab:tll_parameters}. The crossover towards the TLL regime in presence of disorder happens at slightly smaller temperatures than the clean case. This can be explained by the behavior of $\tau_*(T,h\neq 0)$ and $\tau(T,h\neq 0)$: They both decrease with disorder strength, making the dynamical correlation drop much faster than in the clean case, see Fig.~\ref{fig:real_part_vs_time}\,(a). Nonetheless, the parameter-free TLL prediction makes it possible to precisely define the crossover temperature, even in the disordered TLL regime, and this should be a key element to be computed in a more realistically experimental model in order to interpret carefully experimental measurements.

As we cross the superfluid to BG transition at $h_c\simeq 0.37$~\cite{doggen2017}, we observe that on the log-log scale of Fig.~\ref{fig:t1_vs_temp}\,(a), the gradient of the average value of the NMR spin lattice relaxation rate starts to decrease at low temperature as the temperature is lowered, while it increases in the TLL phase. The gradient is plotted in Fig.~\ref{fig:t1_vs_temp}\,(c) where this effects is clearly visible and we believe is a signature of the transition at low temperature. In the TLL phase, the gradient should saturate to $1-1/2K$ as visible for the clean case with $1-1/2K(\Delta=-0.75,h=0)\simeq 0.77$. One would need to access lower temperatures in order to observe the saturation at $h=0.2$ and $0.3$.

In all cases, the average $\overline{1/T_1}$, as previously discussed, is not directly accessible in experiments where a stretched exponential fit to the return-to-equilibrium function $M(t)$, as in Eq.~\eqref{eq:str_exp}, is usually used and the parameter $1/\tau_\mathrm{str}$ interpreted as the ``relevant'' $1/T_1$. Numerically, for each disordered sample, we can compute $M(t)$ and fit its disorder averaged value $\overline{M(t)}$ against a stretched exponential, as it would be done in experiments. The parameters $1/\tau_\mathrm{str}$ and $\theta$ are reported in Fig.~\ref{fig:t1_vs_temp}\,(a) and (b) respectively. Although it is clear that the distributions $P(1/T_1)$ shown in Fig.~\ref{fig:distributions} are not those corresponding to a stretched exponential $\overline{M(t)}$, the value of $1/\tau_\mathrm{str}$ is in very good agreement with the exact average $\overline{1/T_1}$ value, in both the TLL and BG phases. Yet, we can note that a discrepancy seem to appear as we go deeper in the BG phase. For comparison, this is in sharp contrast with results in spin$-1/2$ Heisenberg chains with random exchange couplings, which realize in the low-energy limit the so-called random singlet phase~\cite{shu2018}: At low temperature, it was for instance found that the average $\overline{1/T_1}$ diverges while the stretched exponential estimate was found to go to zero as the temperature was decreased. In that case, the authors found that the stretched exponential did not capture the average value but rather the \textit{typical} value of the NMR spin-lattice relaxation rate, characterized by the median value of the distribution. We report the median value $\mathrm{med}(1/T_1)$ in Fig.~\ref{fig:t1_vs_temp}\,(a) which behaves similarly to the average and the stretched exponential estimates. As thoroughly discussed in Ref.~\onlinecite{johnston2006}, the physical interpretations of the stretching exponent 􏰐$\theta$ are not straightforward, although it is commonly related to the width of the $1/T_1$ distribution. The value of $\theta$ displayed in Fig.~\ref{fig:t1_vs_temp}\,(c) decreases with the disorder strength and seem to decrease with the temperature in the BG phase while it remains roughly constant with temperature in the TLL phase.

\section{Summary and conclusions}\label{sec:summary_conclusions}

Using large-scale simulations based on matrix product state techniques, we have computed the NMR spin-lattice relaxation rate $1/T_1$ in random spin chains displaying a disorder-induced phase transition in the low temperature limit between a critical Tomonaga-Luttinger Liquid phase and a many-body localized phase, known as Bose glass. We have provided numerical evidences that this quantity versus temperature detect the transition and that it should be able to address the still controversial value of the TLL parameter $K$ at criticality, which might be universal~\cite{ristivojevic2012,ristivojevic2014}. One indeed expects at low temperature that $1/T_1\propto T^{1/2K-1}$, where a clear identification of the crossover temperature below which this algebraic dependence becomes valid was possible.

We were also able to access quantities which are not experimentally, such as the $1/T_1$ distributions for different temperatures and disorder strengths in both phases. From this, we discussed the applicability of approximating the disorder averaged return-to-equilibrium function $\overline{M(t)}$ (the only quantity accessible in such NMR experiments) as a stretched exponential $\propto\mathrm{exp}[-(t/\tau_\mathrm{str})^\theta]$. Through this approximation, $1/\tau_\mathrm{str}$ is usually referred as the ``relevant'' NMR spin-lattice relaxation rate but there is no guarantee that it is equal (or simply related) to the actual disorder averaged value of $\overline{1/T_1}$. To emphasize this difference between $\overline{1/T_1}$ and $1/\tau_\mathrm{str}$, it was found that for spin$-1/2$ Heisenberg chains with random exchange couplings~\cite{shu2018}, $\overline{1/T_1}$ diverges at low temperature while the stretched exponential estimate goes to zero. Here, we show that they both behave in a very similar way: A decisive point for experiments to reliably extract $K$ from $1/\tau_\mathrm{str}(T)$.

In the weak disorder regime (not considered in this work, corresponding to the TLL phase close to $\Delta\sim -1/2$), there might be a multiplicative logarithm correction to the $1/T_1$ expression~\eqref{eq:t1_tll} in the TLL phase since the static correlations have also been found to display such a correction in presence of disorder~\cite{ristivojevic2012corr}. This would be qualitatively similar to the one appearing in the better-known isotropic $\Delta=1$ (and clean, $h=0$) case~\cite{barzykin2000,barzykin2001,dupont2016}. However, its effect in presence of disorder would probably be way smaller since it arises with a log at the exponent $2/9$ in the static correlations, very difficult to capture numerically. For instance, there is no mention of including this correction to fit the static correlations in Ref.~\onlinecite{doggen2017}.

It would be interesting to consider next the dynamical spin structure factor $S(q,\omega)$, measured in inelastic neutron scattering experiments. Though it would be much more challenging numerically because involving dynamical correlations at all distances in order to compute the Fourier transform from real to momentum space. In comparison, the study carried out in Ref.~\onlinecite{shu2018} on the dynamical properties of the random singlet phase, was performed in the canonical ensemble (imposing $S^z_\mathrm{tot}=0$), more amenable with the MPS techniques~\cite{nocera2016}. One possibility would be to use quantum Monte Carlo with analytic continuation, where the disorder average could be done before doing the continuation. For the Bose glass phase, the equivalent to the dynamical structure factor in the context of cold atom experiments and Bragg spectroscopy has been investigated at zero temperature in Ref.~\onlinecite{roux2013}, but the nature of the excitations in this phase remains an open question.

\begin{acknowledgments}
    We are grateful to S. Capponi, M. Horvati\'c, N. Laflorencie and J. E. Moore for interesting comments and discussions. This work was funded by the U.S. Department of Energy, Office of Science, Office of Basic Energy Sciences, Materials Sciences and Engineering Division under Contract No. DE-AC02-05-CH11231 through the Scientific Discovery through Advanced Computing (SciDAC) program (KC23DAC Topological and Correlated Matter via Tensor Networks and Quantum Monte Carlo). This research used the Lawrencium computational cluster resource provided by the IT Division at the Lawrence Berkeley National Laboratory (Supported by the Director, Office of Science, Office of Basic Energy Sciences, of the U.S. Department of Energy under Contract No. DE-AC02-05CH11231). This research also used resources of the National Energy Research Scientific Computing Center (NERSC), a U.S. Department of Energy Office of Science User Facility operated under Contract No. DE-AC02-05CH11231. The code for calculations is based on the ITensor library~\footnote{ITensor library, \href{http://itensor.org}{http://itensor.org}.}.
\end{acknowledgments}

\appendix*

\section{Determining the Tomonaga-Luttinger Liquid parameters numerically}

The TLL parameters $u$ and $K$, as well as the prefactor $A$ of the static correlation function $\langle S^\pm_r(0)S^\mp_0(0)\rangle$ at zero temperature have no exact expression at finite disorder strength and have to be determined numerically. The parameters $u$ and $K$ can be related through the hydrodynamic relations to the uniform susceptibility $\chi$ and the spin stiffness $\rho_\mathrm{s}$~\cite{giamarchi2004},
\begin{equation}
    K = \pi\sqrt{\rho_\mathrm{s}\chi}\quad\mathrm{and}\quad u=\sqrt{\rho_\mathrm{s}/\chi}
    \label{eq:tll_parameters_hydro}
\end{equation}
which are both readily computed in quantum Monte Carlo (QMC) through the stochastic series expansion algorithm~\cite{sandvik2002,bauer2011}. We compute these quantities for $N_\mathrm{s}\approx 10^3$ independent random systems of size $L=256$ with periodic boundary conditions at inverse temperature $\beta J=2^{11}$ using the $\beta$-doubling scheme~\cite{sandvik2002bdoubling}. The temperature is sufficiently low to consider that the QMC algorithm is only probing the ground state. Moreover, these quantities do not show strong finite-size effects for the disorder strengths considered here, hence $L=256$ can be taken as the thermodynamic limit result. Similar simulation parameters have been used in Ref.~\onlinecite{doggen2017}, providing very satisfactory results. The average of $\rho_\mathrm{s}$ and $\chi$ over the $N_\mathrm{s}$ samples is then computed to extract the values of $K$ and $u$ according to Eq.~\eqref{eq:tll_parameters_hydro}. The values for $h=0.2$ and $h=0.3$ are reported in Tab.~\ref{tab:tll_parameters} in the main text.

Independently, simulations were also carried out on systems of size $L=256$ with open boundary conditions using the variational density-matrix renormalization group (DMRG) algorithm~\cite{white1992,white1993} at exactly zero temperature, and enforcing $S^z_\mathrm{tot}=0$. Here, the susceptibility and the spin stiffness are not as easily computed but one can compute the transverse static correlation function $\langle S^\pm_i(0)S^\mp_j(0)\rangle$ versus the distance $r=|i-j|$. It has been computed for $N_\mathrm{s}\approx 10^3$ independent samples and the average value has been fitted to the form,
\begin{equation}
    F(i,j,L)=2A\Biggl[\frac{f(i+j,2L)f(i-j, 2L)}{2\sqrt{f(2i,2L)f(2j,2L)}}\Biggr]^{-\frac{1}{2K}}
    \label{eq:tll_corr_fit}
\end{equation}
with $f(i,L)=L|\sin(\pi i/L)|/\pi$. This corresponds to the expected form of $|\langle S^\pm_i(0)S^\mp_j(0)\rangle |$ in a finite system with open boundary conditions, with $i$ and $j$ far enough from the boundaries, and with $|i-j|$ large enough~\cite{cazalilla2004}. In practice, we only consider $|i-L|>L/4$, $|j-L|<3L/4$ with $i<j$ and $|i-j|>10$. The fitted values of $K$ and $A$ are reported in Tab.~\ref{tab:tll_parameters} in the main text for $h=0.2$ and $h=0.3$. The values of $K$ extracted from the QMC and the DMRG agree well with each others.

\input{main.bbl}

\end{document}

%% file: main.bbl
%